\documentclass[iicol,pdflatex,sn-nature]{sn-jnl}

\usepackage{graphicx}%
\usepackage{multirow}%
\usepackage{amsmath,amssymb,amsfonts}%
\usepackage{amsthm}%
\usepackage{mathrsfs}%
\usepackage[title]{appendix}%
\usepackage{xcolor}%
\usepackage{textcomp}%
\usepackage{manyfoot}%
\usepackage{booktabs}%
\usepackage{algorithm}%
\usepackage{algorithmicx}%
\usepackage{algpseudocode}%
\usepackage{listings}%
\usepackage{bibspacing}

\begin{document}

\title[Article Title]{Ultralong-lived Coherent States in Eu$^{3+}$:Y$_2$O$_3$ Optical Ceramics for Quantum Memories}

\author*[1,7]{\fnm{Shuping} \sur{Liu}} \email{liushuping@iqasz.cn}
\author[2,7]{\fnm{Miaomiao} \sur{Ren}}
\author[2,7]{\fnm{Wanting} \sur{Xiao}}
\author[3]{\fnm{Jun} \sur{Wang}}
\author[4]{\fnm{Yuting} \sur{Liu}}
\author[5]{\fnm{Diana} \sur{Serrano}}
\author[5]{\fnm{Philippe} \sur{Goldner}}
\author[6]{\fnm{Dingyuan} \sur{Tang}}
\author[2]{\fnm{Xiantong} \sur{An}}
\author*[2]{\fnm{Fudong} \sur{Wang}} \email{wangfd@sustech.edu.cn}
\author*[1,2]{\fnm{Manjin} \sur{Zhong}} \email{zhongmj@sustech.edu.cn}

\affil[1]{\orgdiv{International Quantum Academy, and Shenzhen Branch, Hefei National Laboratory}, \city{Shenzhen}, \postcode{518048}, \country{China}}
 
\affil[2]{\orgdiv{Shenzhen Institute for Quantum Science and Engineering, Southern University of Science and Technology}, \city{Shenzhen}, \postcode{518055}, \country{China}}

\affil[3]{\orgdiv{School of Physics and Electronic Engineering, Jiangsu Key Laboratory of Advanced Laser Materials and Devices, Jiangsu Normal University}, \city{Xuzhou}, \postcode{221116}, \country{China}}

\affil[4]{\orgdiv{School of Integrated Circuit, Harbin Institute of Technology}, \city{Shenzhen}, \postcode{518055}, \country{China}}

\affil[5]{\orgdiv{Institut de Recherche de Chimie Paris (IRCP), Chimie ParisTech, PSL University, CNRS}, \city{Paris}, \postcode{F-75005}, \country{France}}

\affil[6]{\orgdiv{Future Technology School, Shenzhen Technology University}, \city{Shenzhen}, \postcode{518118}, \country{China}}

\affil[7]{These authors contributed equally to this work.}


\abstract{Rare earth ions (REI) in solid materials are among the leading systems for quantum technology applications. However, developing practical REI quantum devices with long-lived coherent states remains challenging due to great growth difficulties of high-quality REI materials and a lack of comprehensive understanding of REI’s decoherence mechanisms. Here we realize a record optical coherence time of 421.5$\pm$10.5 $\mu$s for the $^7$F$_0\rightarrow^5$D$_0$ transition and more than 30 hours lifetime for the $^7$F$_0$ hyperfine spin states in Eu$^{3+}$:Y$_2$O$_3$ optical ceramics. We report the elimination of two-level-system induced optical decoherence in short-range ordered crystals. Meanwhile, a new decoherence mechanism caused by new kinds of perturbing magnetic centers is identified below 1.5 K. We further demonstrate the coherent light storage over 5 $\mu$s by using the atomic frequency comb protocol. These results open up prospects for the realization of practical quantum memories and large scale quantum communications with REI optical ceramics.}

\keywords{Y$_2$O$_3$, rare earth optical ceramics, optical coherent spectroscopy, decoherence mechanisms, quantum memory}



\maketitle

\section{Introduction}
Quantum technologies typically use two-level superposition states as the fundamental resource (quantum bit, qubit) to realize advanced computing, sensing and communicating functionalities \cite{kimble,Degen,Gisin}. REI are natural qubit carriers with outstanding coherence properties, which enabled by their abundant and less-perturbed 4f-4f optical transitions and optically addressable spin states \cite{Wolfowicz,ZhongManjin}. During recent years, REI doped bulk single crystals with low-magnetic-moment matrix have received considerable attention for developing core quantum devices, such as quantum memories \cite{Liux,ZhongTian}, microwave-to-optical transducers \cite{TXie}, quantum processors \cite{Ruskuc} and single photon sources \cite{Ourari}. Besides, strong efforts have recently been lunched towards versatile REI systems and material topologies, intended for flexible structure design and better integrability. Remarkable progress have been achieved, for instance, in REI optical ceramics, nanoparticles, thin films and molecules in all-optical control of REI nuclear spins \cite{Serranonc,Karlsson}, fast modulation of REI emissions \cite {Casabone}, efficient optical spin initialization \cite {Kumar,Diana}, nanoscale frequency-multiplexed quantum storage \cite{Fossati} and even REI-diamond hybrid quantum system designs \cite{Balasa}.

Nevertheless, a challenge to develop REI systems for practical quantum technology application is that the coherence properties of REI are commonly deteriorated by perturbations from local environments, whose nature are not comprehensively understood. Notably, interactions of REI with dynamical disorder modes, so called two-level-system (TLS) have been observed in many REI doped single crystals \cite{Flinn, Macfarlane3, Macfarlane5, Thiel}, and ascribed to be the dominant decoherence mechanisms in REI doped nanoparticles and ceramics \cite{Kunkel,Macfarlane4}. As yet, the nature and origin of the TLS are not completely identified. On the other hand, it remains a difficult problem to control the fabrication of REI materials with less environment fluctuations and superior coherence properties, due to a lack of clear understanding of the correlation between the microstructure and the decoherence processes. The growth and reproducibility of high-quality REI single crystals are also impeded by the ultrahigh melting temperatures of many host matrix. 


In this work, we demonstrate a record optical coherence time of 421.5$\pm$10.5 $\mu$ as well as an elimination of TLS decoherence dynamics in Eu$^{3+}$:Y$_2$O$_3$ optical ceramics, which fabricated at a temperature far below Y$_2$O$_3$'s melting point (T$_m$=2400 $^{\circ}\mathrm{C}$). We also identify a new decoherence mechanism in ultra-low temperature regime. By comparing the microstructure, low temperature EPR and the optical coherence property of the ceramics, we clarify the role of different kinds of defects in the TLS and the decoherence process of Eu$^{3+}$ ions. A superior ground hyperfine state lifetime of more than 30 h and an atomic frequency comb memory with 5 $\mu$s storage time are also demonstrated. This study would trigger further developments of REI optical ceramics for practical quantum memory applications, and provide a guidance to improve the coherence properties of a variety of REI systems by means of chemical and physical manipulations.

\section{Results}
\textbf{Microstructure and perturbing center characterizations.}

\begin{figure*}[h]
	\centering
	\includegraphics[width=0.9\textwidth]{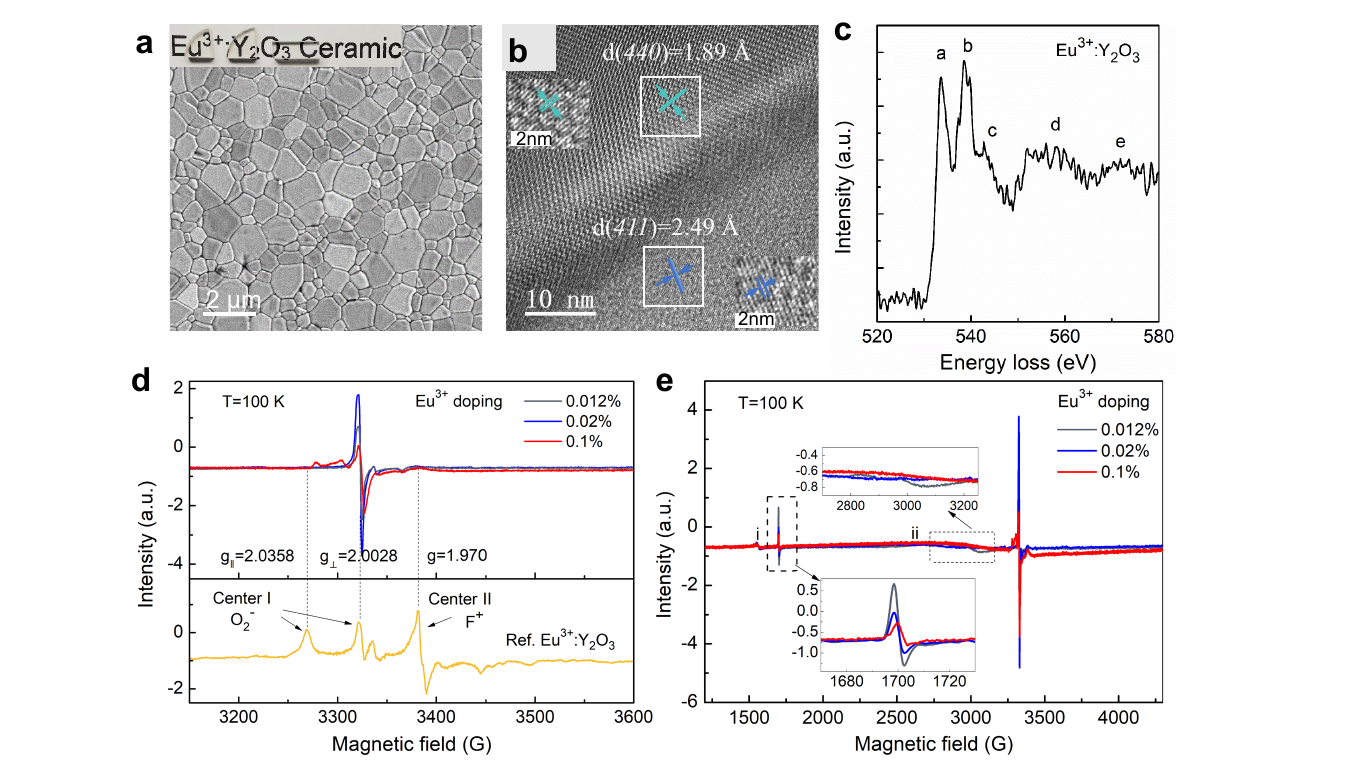}
	\caption{Microstructure and perturbing centers in Eu$^{3+}$:Y$_2$O$_3$ ceramics. a) Photographs of the ceramics with different Eu$^{3+}$ concentrations and the SEM images of representative sample (0.012\% Eu$^{3+}$). b) HRTEM image of the ceramic shows good crystalline structure of the grain boundaries. c) The EELS of oxygen K edge after background subtraction near the grain boundary of the sample. d) low temperature EPR spectra of oxygen related point defects in the matrix. The spectrum of 0.3\% Eu$^{3+}$:Y$_2$O$_3$ nanoparticles is given as a reference, in which two typical signals are identified: center I interstitial superoxide anions O${_2^-}$ (g$_\Vert$=2.0358, g$_\bot$=2.0028) and center II charged oxygen vacancies F$^+$(g=1.976), respectively. e) Broad band scanning of the EPR spectra indicates other weaker defects (feature i and ii) with unpaired electrons. }\label{fig1}
\end{figure*}

The Eu$^{3+}$:Y$_2$O$_3$ optical ceramics we studied were fabricated by vacuum sintering following by hot-isostatic pressing (HIP) processes and post-annealing treatment, in which the Y$_2$O$_3$ nano-powders were synthesized by co-precipitation method from 99.99\% raw materials (See Methods and Supplementary section 1). As Fig. 1a and Supplementary Fig. S1 depicts, even without using any sintering aids, the 0.012\%, 0.02\% and 0.1\% Eu$^{3+}$ doped ceramics show high optical qualities as well as very dense microstructures. No secondary phase is observed both inside of the grains and at grain boundaries, consistent with the pure Y$_2$O$_3$ cubic phase verified by powder X-ray diffraction (XRD) measurement (Supplementary Fig. S2). The high-resolution transmission electron microscopy (HRTEM) image in Fig. 1b shows good crystalline structure of the grain boundary area, where clear and clean boundary with a thickness within 10 nm is observed (see more HRTEM images and analysis in Supplementary Fig. S3). Besides, the electron energy loss spectroscopy (EELS) of oxygen \emph{K} edge of the ceramic (Fig. 1c) displays a similar shape with the stoichiometric Y$_2$O$_3$ \cite{Travlos}. This suggests that a 10-hour-long air-annealing treatment can cure effectively oxygen vacancies, which are considered one of the main perturbing sources in the matrix \cite{Kunkel}. 

Electron paramagnetic resonance (EPR) measurements reveal also a rather small amount of oxygen interstitial defects in our optical ceramics. As shown in Supplementary section 2 and Fig. S4, both F$^+$ centers (oxygen vacancies with one electron left, g=1.970) and O${_2^-}$ (oxygen interstitials, g$_\vert$=2.0358, g$_\bot$=2.0028) \cite{Singh} are hardly observed in our ceramics at room temperature (RT), which is distinct from previous reports \cite{Kunkel3}. At 100 K, one can observe only the g$_\bot$ signal of O${_2^-}$ centers clearly (Fig. 1d) when compared with the reference Eu$^{3+}$:Y$_2$O$_3$ nanoparticles. It is worth noting that the 0.1\% Eu$^{3+}$ sample exhibits a much lower O${_2^-}$ intensity as well as two additional weaker peaks at about 3300 G. This is attributed to the different fabrication procedures used for this sample. Besides, weaker EPR signals of other point defects are also observed at lower magnetic fields, as shown in Fig. 1e. This may come from impurities that introduced from raw materials or fabrication process.

\begin{figure*}
	\centering
	\includegraphics [width=0.9\textwidth] {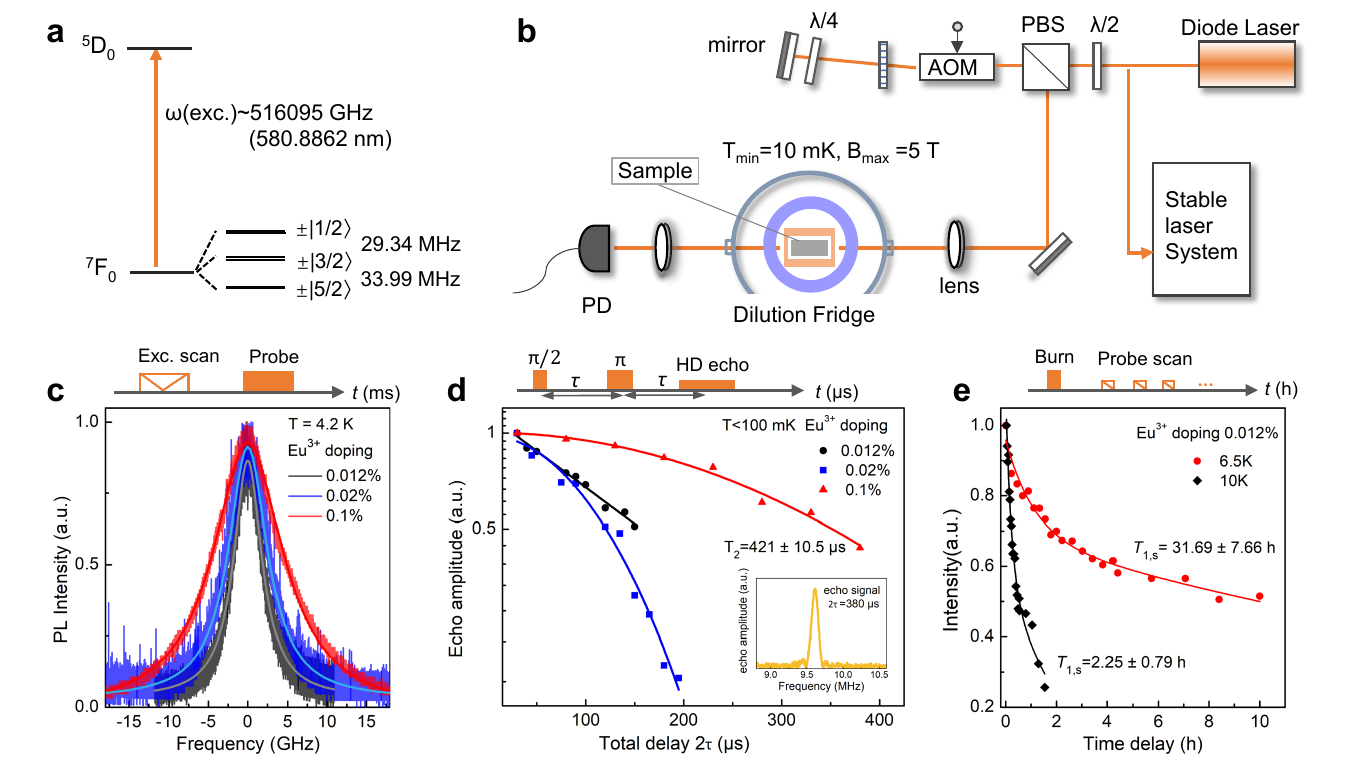}
	\caption{High resolution and coherent optical spectroscopic experiments. a) $^7$F$_0 \rightarrow ^5$D$_0$ optical transition and $^7$F$_0$ ground-state hyperfine structure of $^{151}$Eu$^{3+}$ in Y$_2$O$_3$. b) Experimental design and set-up. c) Inhomogeneous linewidth of the optical transition measured at 4.2 K. Solid lines: Lorentzian fit corresponding to a full width at half maximum of 4.34 GHz, 6.03 GHz and 10.76 GHz for the 0.012\%,0.02\% and 0.1\% Eu$^{3+}$ doping samples, respectively. The transition frequency at 0 detuning corresponds to 516095 $\pm$ 1 GHz (580.8862 $\pm$ 0.0001 nm) for the sample. d) 2PPE decays measured below 100 mK for Eu$^{3+}$:Y$_2$O$_3$ optical ceramics with different Eu$^{3+}$ concentrations. The optical $T_2$ values are derived by fitting the echo decay curves according to A=A$_0$ exp[-(2$\tau/T_2)^x$] (solid line). e) Spectral hole area decays measured at 6.5 K and 10 K, respectively for the 0.012\% ceramic. Solid lines: double exponential fits.} 
	\label{Figure 2}
\end{figure*}

\textbf{High-resolution and coherent optical spectroscopy.}
The RT photoluminescence (PL) spectra of Eu$^{3+}$:Y$_2$O$_3$ ceramics are shown in Supplementary Fig. S5. Among the $^5$D$_J (J=0, 1, 2)\rightarrow^7$F$_J$ (J=0, 1, 2,3) optical transitions, the weak $^5$D$_0\rightarrow^7$F$_0$ transition at about 580 nm is the least affected by magnetic field fluctuations of the local environments, making it preferred for quantum technologies. Therefore, the following experiments performed from about 40 mK to 11 K are mainly focused on this transition. The relevant energy levels and  ground-state hyperfine structure of $^{151}$Eu$^{3+}$ in Y$_2$O$_3$ are shown in Fig. 2a. Experimental setups, including a highly stable laser system and a dilution fridge, are shown in Fig. 2b and detailed in Supplementary Fig. S6.

At 4.2 K, the $^7$F$_0\rightarrow^5$D$_0$ transition of the ensemble Eu$^{3+}$ ions occurs at 580.8862 nm ($\omega$=516095 GHz) in vacuum and exhibits a much narrower inhomogeneous linewidth (Fig. 2c). This is ascribed to the suppression of two-phonon Raman processes \cite{Eloïse}. A Lorentzian function can fit well the lineshape, giving a full width at half maximum (FWHM, $\Gamma_\mathrm{inh}$) of 4.34 GHz, 6.03 GHz and 10.76 GHz for the 0.012\%, 0.02\% and 0.1\% Eu$^{3+}$ sample, respectively. The former value is more than 1 GHz narrower than 0.004\% Eu$^{3+}$:Y$_2$O$_3$ single crystal ($\Gamma_\mathrm{inh}$=5.5 GHz) \cite{Flinn}, indicating a good suppression of static perturbations in our ceramics. The larger $\Gamma_\mathrm{inh}$ observed in the 0.1\% and 0.02\% Eu$^{3+}$ ceramics are attributed to the higher doping level and  larger amount of O${_2^-}$ centers that detected by EPR measurements (Fig. 1d), respectively.

\begin{figure*}
	\centering
	\includegraphics [width=0.9\textwidth]{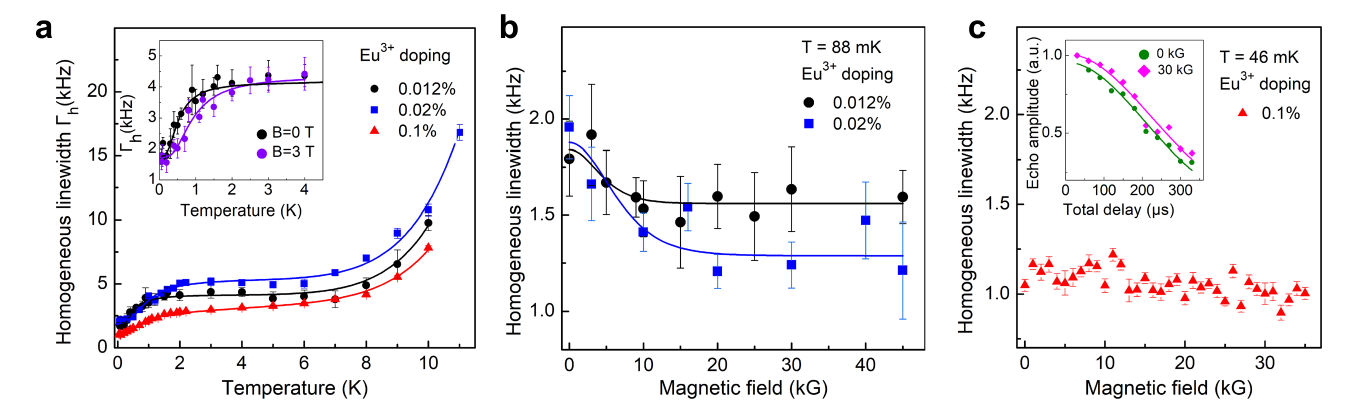}
	\caption{Optical decoherence mechanism studies. a) Temperature dependence of optical $\Gamma_\mathrm{h}$ for Eu$^{3+}$:Y$_2$O$_3$ optical ceramics with different Eu$^{3+}$ concentrations. In the inset is the zoomed $\Gamma_\mathrm{h}$ of the 0.012\% Eu$^{3+}$ sample in 40 mK-4.5 K region measured with and without a magnetic field of 3 T. Solid lines: fits to the data using Eq. (1). b) Magnetic field dependence of $\Gamma_\mathrm{h}$ for the 0.012\%, 0.02\% Eu$^{3+}$ samples measured at 88 mK. Solid lines are fits to the data using the spectral diffusion model in Eq. (1). c) Magnetic field dependence of $\Gamma_\mathrm{h}$ for the 0.1\% Eu$^{3+}$ sample measured at 45 mK. In the inset shows the echo decay curves measured with and without magnetic field.}
	\label{figure3}
\end{figure*} 

Compared to inhomogeneous broadening, optical homogeneous linewidth $\Gamma_\mathrm{h}$ ($\Gamma_\mathrm{h}$=1/($\pi T_2$)), or optical coherence time $T_2$ is of key importance for quantum technologies and is extremely sensitive to dynamic perturbations, such as phonons, electric or magnetic field fluctuations and TLS. As displayed in Fig. 2d, the optical $T_2$ of Eu$^{3+}$, measured by two-pulse photon echo technique below 100 mK (2PPE, see Methods and Supplementary Fig. S7), is observed the longest in the 0.1\% Eu$^{3+}$ sample which possesses the largest optical $\Gamma\mathrm{_{inh}}$ but the lowest O${_2^-}$ density. As a contrast, the shortest $T_2$ is obtained in the one contents the highest O${_2^-}$ density (Fig. 1d). This reveals that the decoherence effect induced by O${_2^-}$ centers is more severe than the instantaneous spectral diffusion (ISD) effect induced by Eu$^{3+}$ themselves \cite{Huang} and other point defects observed in Fig. 1e. It is worth noting that the optical $T_2$ we measured in 0.012\%, 0.02\% and 0.1\% Eu$^{3+}$ ceramic is 190.2 $\pm$ 8.7 $\mu$s, 145.7 $\pm$ 4.7 $\mu$s and 421.5 $\pm$ 10.5 $\mu$s, respectively. Such values are 2-10 times longer than that reported in previous optical ceramics and nanoparticles \cite{Bartholomew,Kunkel,Shuping}. The optical $T_2$ of 0.1\% Eu$^{3+}$ ceramic even compares favorably with that of the best Eu$^{3+}$:Y$_2$O$_3$ single crystals ever reported ($T_2$ = 130-420 $\mu$s)\cite{Macfarlane}. Moreover, the optical coherence performance of our ceramics display a minimal spectral diffusion of subkilohertz over 1 ms time scale (Supplementary Fig. S8), indicating the high quality of the optical ceramics.

The population lifetime of $^7$F$_0$ ground hyperfine states $T_{1,s}$ are also measured on the 0.012\% Eu$^{3+}$ ceramic using hole burning technique (Supplementary Fig. S9) \cite{Goldner}. The results shown in Fig. 2e suggest that the longer component of $T_{1,s}$ at 6.5 K is 31.69±7.66 h, which is more than 2 orders of magnitude longer than that obtained in previous optical ceramics ($T_{1,s}$=15±5 min at 6 K) \cite{Kunkel2} and also comparable to that of the best Eu$^{3+}$:Y$_2$O$_3$ single crystal \cite{Babbitt}. This result indicates that the hyperfine spin state lifetime of Eu$^{3+}$ is not limited by the nature of polycrystalline ceramics. Therefore a ultralong theoretical spin $T_2$ of over 2 days ($T_{\rm 2,s}$=2$T_{\rm 1,s}$) in Eu$^{3+}$:Y$_2$O$_3$ ceramics is predicted, which enables superior quantum storage performance.

\textbf{Optical decoherence mechanisms.} The temperature dependence of $\Gamma_\mathrm{h}$ which measured at 40 mK-11 K without magnetic field is shown in Fig. 3a. It is seen that the $\Gamma_\mathrm{h}$ values of all the ceramics start to increase rapidly at T $>$ 8 K. This is consistent with the two-phonon Raman (TPR) scattering process ($T_2$ $\propto$ T$^7$) \cite{Kunkel}. At temperatures below 8 K, it was reported that the $\Gamma_\mathrm{h}$ was limited by TLS in most of amorphous and less-ordered crystals. Such mechanism typically results in a nearly linear change of $\Gamma_\mathrm{h}$ along temperatures ($T_2$ $\propto$ T)\cite{Bartholomew,Flinn}. However, the TLS effect is not observed in the 0.012\% and 0.02\% Eu$^{3+}$ samples,
and is observed very weak in the 0.1\% Eu$^{3+}$ one. Notably, when temperature decreases from about 1.5 K down to subkelvin region, a dramatic non-linear decrease of $\Gamma_\mathrm{h}$ is observed in all samples. Such decoherence dynamics still exist after applying an external magnetic field of 3 T during measurements, as shown in the inset of Fig. 3a. To the best of our knowledge, this phenomenon is observed for the first time in Eu$^{3+}$:Y$_2$O$_3$ either in single crystal or in polycrystalline systems. 

Herein, we deduce that there exist some other sources of Eu$^{3+}$ decoherence that are not identified before. By considering the TPR, TLS and a spectral diffusion model caused by possible perturbing centers \cite{Böttger}: 
\begin{equation}
	\Gamma\mathrm{_h}(T)=\Gamma\mathrm{_0} + A\,{\rm sech^2}({\frac{\xi}{2k_{\rm B}T}})+ \alpha_{\rm TLS}\,T + \alpha_{\rm TPR}\,T^7 \label{eq3}
\end{equation}
where $\Gamma\mathrm{_0}$ is the residual homogeneous broadening and $\xi$ the parameter given by the energy splitting of the perturbing centers in the environment. $\xi$ can also be described by the energy difference between two states $\xi=g_{\rm env}\mu_{\rm B} B$, where $g_{\rm env}$ is the effective g-factor of the randomly oriented perturbing centers, $\mu_{\rm B}$ the Bohr magneton, $B$ the magnetic field. $T$ is the temperature, $k_{\rm B}$ is the Boltzmann constant, $\alpha_{\rm TLS}$ and $\alpha_{\rm TPR}$ are the TLS and TPR coupling constant, respectively. 

\begin{figure*}
	\centering
	\includegraphics [width=0.9\textwidth]{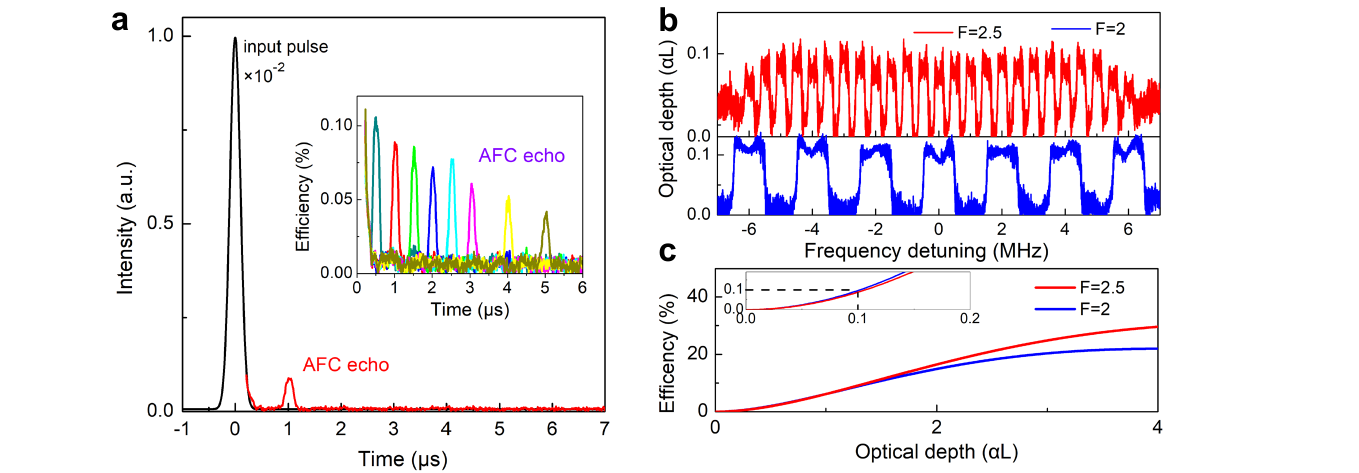}
	\caption{Coherent light storage demonstrations in Eu$^{3+}$:Y$_2$O$_3$ optical ceramics. a) The storage process: the black line is the input pulse (intensity $\times$ 10$^{-2}$) and the red line is storage echo of 1 $\mu$s. In the insert are AFC echoes with variable storage times obtained by changing AFC tooth separations. The longest storage time reaches 5 $\mu$s and the output pulse at 0.5 $\mu$s with a storage efficiency of 0.1\% in agree with theory. b) The atomic frequency comb (AFC) created with different comb finesses by optical pumping within the inhomogeneous line of $^7$F$_0\rightarrow^5$D$_0$ transition of Eu$^{3+}$ ions. c) Theoretical storage efficiencies as a function of optical depth }
	\label{figure4}
\end{figure*}

The fit, shown as the solid line in Fig. 3a, gives good agreement with experimental results. The best-fit parameters to Eq. (1) are given in Supplementary Table S1.
It is seen that $\alpha_{\rm TLS}$=0 for both the 0.012\% and 0.02\% Eu$^{3+}$ samples, and is 0.14 kHz/K for the 0.1\% Eu$^{3+}$ ceramic. The later value, nevertheless, is more than 4 times smaller than that of previous Eu$^{3+}$:Y$_2$O$_3$ optical ceramics \cite{Kunkel}, and 2 orders of magnitude smaller than that of relevant nanoparticles \cite{Fossati2}. This confirms the efficient suppression and even elimination of the TLS effect in our ceramics. Besides, the presence of a new kind of perturbing center is evidenced, whose energy splitting is characterized by $\xi/2k_{\rm B}$=0.47-0.9 K without magnetic field. Under $B$=3 T, the energy splitting for the 0.012\% Eu$^{3+}$ sample is observed enlarged from 0.47 K to 0.85 K. This indicates the paramagnetic nature of the perturbing centers. Calculation from $\xi=g_{env}\mu_{\rm B} B$ gives an effective $g_{\rm env}$ factor of 0.84. 

At subkelvin temperatures, the $\Gamma\mathrm{_h}$ exhibits a small further decrease with external magnetic fields for the 0.012\% and 0.02\% Eu$^{3+}$ samples (Fig. 3b) and very little variation for the 0.1\% Eu$^{3+}$ one (Fig. 3c). Though the data of the former two present higher experimental incertitude, a fit is made to the same spectral diffusion model described in Eq. (1) and shown as the solid line in Fig. 3b,
\begin{equation}
	\Gamma\mathrm{_h}(B)=\Gamma\mathrm{_0} + A\,{\rm sech^2}({\frac{g_{\rm env}\mu_{\rm B} B}{2k_{\rm B}T}}) \label{eq2}
\end{equation}

where $\Gamma\mathrm{_0}$, $g_{\rm env}$, $\mu_{\rm B}$, $B$, $T$, and $k_{\rm B}$ the same meaning in Eq (1). This fit gives $g_{\rm env}$=0.35-0.51 (Supplementary Table S2), which is consistent with that derived from Fig. 3a. Therefore, we conclude that perturbations from the new kind of perturbing centers can be significantly suppressed by decreasing temperatures to subkelvin level, and be further suppressed by applying an external magnetic field.

\textbf{Coherent light storage demonstration} 
As shown in Fig. 4, we finally demonstrate coherent optical storage in Eu$^{3+}$:Y$_2$O$_3$ optical ceramics by using atomic frequency comb (AFC) memory protocol \cite{Afzelius} (See details in Supplementary section 4). Fig. 4b presents the periodic frequency comb structure spanning about 12 MHz range within the inhomogeneous line of the ceramics. By optimizing the width of comb teeth and the finesse $F$ of frequency comb, the output pulses are observed up to 5 $\mu$s storage time. This is 400 \% longer than that of the recently reported Eu$^{3+}$ molecular crystal \cite{Diana}. The storage efficiency $\eta$ is about 0.11 \% with a storage time $t_s$ of 0.5 $\mu$s, in consistent with theoretical calculations that shown in Fig. 4c. Further storage time and efficiency enhancement can be achieved by increasing the $F$ of the comb teeth with a suitable laser, transferring the coherent signal to ground hyperfine levels, and increasing the optical depth of the sample or coupling it to an optical cavity, respectively.

\section{Discussion}
The longest optical $T_2$ we achieved for the Eu$^{3+}$:Y$_2$O$_3$ optical ceramics is 421.5 $\pm$ 10.5 $\mu$s, which is about 2-10 times longer than the previous optical ceramics and nanoparticles \cite{Kunkel,Shuping}. The $^7$F$_0$ hyperfine spin states lifetime is obtained as 31.69±7.66 h at 6.5 K, which is more than 2 orders of magnitude longer than that in previous reports. Both the optical $T_2$ and the $^7$F$_0$ hyperfine spin states lifetime compare favorably with the best Eu$^{3+}$:Y$_2$O$_3$ single crystals ever reported, revealing that they are not limited by the polycrystalline nature of optical ceramics. Besides, we show successful coherent light storage in such optical ceramics with 5 $\mu$s storage time and theoretical storage efficiency. Since the fabrication of optical ceramics is much more technically feasible, environmental friendly, time-saving and financially efficient. Our work, therefore, demonstrates the tremendous potential of REI optical ceramics to develop practical quantum memory devices for quantum communication networks.

From the temperature dependence of the $\Gamma\mathrm{_h}$, we see the elimination of TLS effect for the first time in the 0.012\% and 0.02\% Eu$^{3+}$ ceramics, in which higher density of O${_2^-}$ centers as well as shorter optical $T_2$ are found when compared to the 0.1\% Eu$^{3+}$ sample. This suggests that the TLS may be not dominated by the oxygen related point defects in the samples, which contrasts previous reports \cite{Perrot, Kunkel}. Based on the microstructure, low-temperature EPR, and optical decoherence dynamic studies, we propose that the TLS is most possibly caused by high-dimensional defects, such as dislocations and/or grain boundaries in the microstructure. This is reasonable since a larger number of Eu$^{3+}$ ions which located near grain boundaries or dislocations are being excited in the 0.1\% Eu$^{3+}$ ceramic, taking into account the higher Eu$^{3+}$ concentration and longer optical pass length of this sample. This statement can also explain the decrease of $\alpha_{\rm TLS}$ and $\Gamma\mathrm{_h}$ that recently reported on oxygen plasma treated Eu$^{3+}$:Y$_2$O$_3$ nanoparticles \cite{Fossati2}, in which more F$^+$ point defects were induced and high-dimensional defects were healed by the treatment \cite{Shuping,Wuss}.

A new decoherence mechanism is identified at temperatures below 1.5 K. This is found to be caused by perturbing magnetic centers with an estimated g-factor of 0.84. This value is much smaller than that of oxygen related defects (O${_2^-}$ and F$^+$ centers) and other magnetic impurities (g$>$2), but the related signals are not detected by low temperature EPR measurements. As a result, we deduce that this kind of perturbing magnetic centers may come from trace impurities in raw materials, or localized low energy phonon modes in the microstructures \cite{Macfarlane4}. Further study is still needed to clarify the origin of them. Nevertheless, the findings about the decoherence mechanisms in this work will be highly instructive not only in optimizing the coherence performance of various existing REI materials, but also in exploring novel REI quantum systems with tailorable properties.

\section{Methods}

\textbf{High performance optical ceramic fabrication.} The Eu$^{3+}$:Y$_2$O$_3$ optical ceramics were fabricated through co-precipitation, ball milling, vacuum sintering without sintering aids, hot isostatically pressing (HIP) and subsequently post air-annealing processes (detailed in Supplementary section 1). After being polished and cut, the fabricted 0.012\% and 0.02\% Eu$^{3+}$ ceramics with an optical length of 4 mm and the 0.1\% Eu$^{3+}$ one with 14 mm were used for high resolution spectroscopic experiments. The length difference would not influence the optical coherence properties of the samples. While a stronger echo signal and broader inhomogenous linewidth are detected from a longer sample due to the increased amount of Eu$^{3+}$ ions and other negative factors that the laser pulse could addressed.   

\textbf{Microstructure and perturbing center characterizations.} The microstructure of the ceramics was observed with a scanning electron microscope (SEM, zeiss). Local structures nearby grain boundaries and the electron energy loss spectrum (EELS) were measured by high resolution transmission electron microscope (HRTEM, TECNAI G2F30, Thermo Fisher) which equipped with a GATAN-PEELS spectrometer. Electric paramagnetic resonance (EPR) experiments were carried out at 100 K by using a Bruker EMXPlus-10/12 spectrometer and at 4.5 K, 20 K, 50 K, RT by using a CIQTEK EPR100 spectrometer. Both of them were operated at X band (9.5 GHz), and the same amount of grinded powders were used for each sample during each measurements. 

\textbf{cryogenic temperature, high-resolution and coherent optical spectroscopy.} The ceramics were placed on a homebuilt mount in a dilution fridge (LD400, Bluefors), whose lowest temperature reaches to $\sim$10 mK without load. A high precision three-dimensional vector superconducting magnet (American Magnetics Inc.) was equipped to provide external magnetic field up to B=5 T in Z axis. A highly stabilized diode laser (Toptica, 10 kHz linewidth) was used to initialize and excite the $^7$F$_0\rightarrow^5$D$_0$ optical transition of $^{151}$Eu$^{3+}$ ions in $C_2$ symmetry site of Y$_2$O$_3$ at 580.8862 nm (516095 GHz). The pulse sequences was gated with a double pass acousto-optic modulator (AOM) which is driven by an arbitrary waveform generator(AWG). Due to the polycrystalline character of optical ceramics, the scattered signals from the backside of the samples were collected by a series of lenses and finally detected by an APD or a PMT detector, followed with suitable filters. 

The inhomogeneous linewidth, AFC optical memory performance, optical $T_2$, ground hyperfine state lifetime $T_{1,s}$ were measured by using optical pumping, 2PPE and hole-burning techniques, respectively. The related laser pulse sequences and detailed parameters used in this work were detailed in Supplementary section 3.

\bmhead{Data availability} 
The data that support the findings of this study are available from the corresponding author on request.

\bibliography{sn-Y2O3}

\bmhead{Acknowledgements}
We thank Dr. Wang Jun's team in fabrication and optimization of the high quality Eu$^{3+}$:Y$_2$O$_3$ transparent ceramics and Dr. Kangwei Xia in EPR measurements. This work was supported by the Innovation Program for Quantum Science and Technology (No. 2021ZD0301204), the National Natural Science Foundation of China (Grant Nos. 12304454, 62105130, 11904159 and 12004168), National Key Research and Development Program of China (Grant No. 2022YFB3605800), Guangdong Basic and Applied Basic Research Foundation (Grant No. 2021A1515110191), Guangdong Innovative and Entrepreneurial Research Team Program (Grant No. 2019ZT08X324), the Key-Area Research and Development Program of Guangdong Province (Grant No. 2018B030326001) and The Science, Technology and Innovation Commission of Shenzhen Municipality (KQTD202108110900-49034).


\bmhead{Supplementary information}
Supplementary Information accompanies this paper is available.

\bmhead{Competing interests}
The authors declare no competing interests.



\end{document}